\documentclass[aps,prd,unsortedaddress,superscriptaddress,showpacs,twocolumn]{revtex4-1}
\usepackage{verbatim}
\usepackage{graphics}
\usepackage{graphicx}
\usepackage{epstopdf}
\usepackage{epsfig}
\usepackage{color}
\usepackage{dcolumn}

\begin{document}
\title{A search for two body muon decay signals}

%\affiliation{University of Alberta, Edmonton, AB, T6G 2J1, Canada}
%\affiliation{University of British Columbia, Vancouver, BC, V6T 1Z1, Canada}
%\affiliation{Kurchatov Institute, Moscow, 123182, Russia}
%\affiliation{University of Montreal, Montreal, QC, H3C 3J7, Canada}
%\affiliation{University of Regina, Regina, SK, S4S 0A2, Canada}
%\affiliation{Texas A\&M University, College Station, TX 77843, U.S.A.}
%\affiliation{TRIUMF, Vancouver, BC, V6T 2A3, Canada}
%\affiliation{Valparaiso University, Valparaiso, IN 46383, U.S.A.}

\author{R.~Bayes}\thanks{Corresponding author: Ryan.Bayes@glasgow.ac.uk}
\altaffiliation[Present Address: ]{Univ.\@ of Glasgow,
Glasgow, G12 8QQ, United Kingdom}\altaffiliation[Affiliated with: ]{Univ.\@ of Victoria,
Victoria, British Columbia.}
\affiliation{TRIUMF, Vancouver, British Columbia, V6T 2A3, Canada}

\author{J.F.~Bueno}
\affiliation{University of British Columbia, Vancouver, British Columbia, V6T 1Z1, Canada}

\author{Yu.I.~Davydov}
\altaffiliation[Present Address: ]{JINR,
Dubna, Russia.}
\affiliation{TRIUMF, Vancouver, British Columbia, V6T 2A3, Canada}

\author{P.~Depommier}
\affiliation{University of Montreal, Montreal, Quebec, H3C 3J7, Canada}

\author{W.~Faszer}
\affiliation{TRIUMF, Vancouver, British Columbia, V6T 2A3, Canada}

\author{M.C.~Fujiwara}
\affiliation{TRIUMF, Vancouver, British Columbia, V6T 2A3, Canada}

\author{C.A.~Gagliardi}
\affiliation{Texas A\&M University, College Station, TX 77843, U.S.A.}

\author{A.~Gaponenko}
\altaffiliation[Present address: ]{FNAL, Batavia, Illinois, USA.}
\affiliation{University of Alberta, Edmonton, Alberta, T6G 2J1, Canada}

\author{D.R.~Gill}
\affiliation{TRIUMF, Vancouver, British Columbia, V6T 2A3, Canada}

\author{A.~Grossheim}
\affiliation{TRIUMF, Vancouver, British Columbia, V6T 2A3, Canada}

\author{P.~Gumplinger}
\affiliation{TRIUMF, Vancouver, British Columbia, V6T 2A3, Canada}

\author{M.D.~Hasinoff}
\affiliation{University of British Columbia, Vancouver, British Columbia, V6T 1Z1, Canada}

\author{R.S.~Henderson}
\affiliation{TRIUMF, Vancouver, British Columbia, V6T 2A3, Canada}

\author{A.~Hillairet}
\altaffiliation[Affiliated with: ]{Univ.\@ of Victoria,
Victoria, British Columbia.}
\affiliation{TRIUMF, Vancouver, British Columbia, V6T 2A3, Canada}

\author{J.~Hu}
\altaffiliation[Present address: ]{AECL, Mississauga, ON, Canada, L5K 1B2}
\affiliation{TRIUMF, Vancouver, British Columbia, V6T 2A3, Canada}

\author{D.D.~Koetke}
\affiliation{Valparaiso University, Valparaiso, Indiana 46383, U.S.A.}

\author{R.P.~MacDonald}
\affiliation{University of Alberta, Edmonton, Alberta, T6G 2J1, Canada}

\author{G.M.~Marshall}
\affiliation{TRIUMF, Vancouver, British Columbia, V6T 2A3, Canada}

\author{E.L.~Mathie}
\affiliation{University of Regina, Regina, Saskatchewan, S4S 0A2, Canada}

\author{R.E.~Mischke}
\affiliation{TRIUMF, Vancouver, British Columbia, V6T 2A3, Canada}

\author{K.~Olchanski}
\affiliation{TRIUMF, Vancouver, British Columbia, V6T 2A3, Canada}

\author{A.~Olin}
\altaffiliation[Affiliated with: ]{Univ.\@ of Victoria,
Victoria, British Columbia.}
\affiliation{TRIUMF, Vancouver, British Columbia, V6T 2A3, Canada}

\author{R.~Openshaw}
\affiliation{TRIUMF, Vancouver, British Columbia, V6T 2A3, Canada}

\author{J.-M.~Poutissou}
\affiliation{TRIUMF, Vancouver, British Columbia, V6T 2A3, Canada}

\author{R.~Poutissou}
\affiliation{TRIUMF, Vancouver, British Columbia, V6T 2A3, Canada}

\author{V.~Selivanov}
\affiliation{Kurchatov Institute, Moscow, 123182, Russia}

\author{G.~Sheffer}
\affiliation{TRIUMF, Vancouver, British Columbia, V6T 2A3, Canada}

\author{B.~Shin}
\altaffiliation[Affiliated with: ]{Univ.\@ of Saskatchewan,
Saskatoon, SK.}
\affiliation{TRIUMF, Vancouver, British Columbia, V6T 2A3, Canada}

\author{T.D.S.~Stanislaus}
\affiliation{Valparaiso University, Valparaiso, Indiana 46383, U.S.A.}

\author{R.~Tacik}
\affiliation{University of Regina, Regina, Saskatchewan, S4S 0A2, Canada}

\author{R.E.~Tribble}
\affiliation{Texas A\&M University, College Station, TX 77843, U.S.A.}

\collaboration{TWIST Collaboration}
\noaffiliation

\date{\today}

\begin{abstract}
  Lepton family number violation is tested by searching for $\mu^+\to
  e^+X^0$ decays among the 5.8$\times 10^8$ positive muon decay events
  analyzed by the TWIST collaboration. Limits are set on the
  production of both massless and massive $X^0$ bosons. The large
  angular acceptance of this experiment allows limits to be placed on
  anisotropic $\mu^+\to e^+X^0$ decays, which can arise from
  interactions violating both lepton flavor and parity
  conservation. Branching ratio limits of order $10^{-5}$\ are
  obtained for bosons with masses of 13 - 80 MeV/c$^2$ and with different
  decay asymmetries. For bosons with
  masses less than 13 MeV/c$^{2}$ the asymmetry dependence is much
  stronger and the 90\% limit on the branching ratio varies up to $5.8
  \times 10^{-5}$.  This is the first study that explicitly evaluates
  the limits for anisotropic two body muon decays.
\end{abstract}
\maketitle

\section{Introduction} %  > A discussion of the motivation and history

The conservation of lepton family number, or flavor, in reactions
involving charged leptons is a postulate of the standard model
(SM). Positive muon decay ($\mu^+\to e^+\nu_e\bar{\nu}_{\mu}$) is an excellent
low energy system with which to search for charged lepton flavor
violating (CLFV) interactions, such as the decay to a positron and an
unknown neutral boson $\mu^+ \to e^+X^0$, for the same reasons that it
is attractive for weak interaction tests: muons can be produced in
large quantities and the decay is observed with very low backgrounds.

Early studies of muon decay rejected a two body final state as the
normal decay~\cite{Hincks:1948vr}, an unexpected result at the
time. When the final state of the CLFV decay products can be detected,
very stringent exclusive limits have been placed on the branching
ratio of the decay. This is the case for the detection of $\mu^+\to
e^+\gamma$~\cite{Adam:2011ch,Forero:2011pc,Adam:2013mnn} or $\mu^+\to
e^+X^{0}, X^{0}\to e^+e^-$~\cite{Eichler:1986nj} processes. However,
if the neutral $X^{0}$\ boson or its decay products are not detected,
only the shape of the positron spectrum is available to set an
inclusive limit on the decay process.

Stable, non-interacting $X^0$ bosons have been associated with
particles such as axions~\cite{Wilczek:1982rv} and
Majorons~\cite{Chikashige:1980ui,Aulakh:1982yn}.  The $X^{0}$ boson is
massless when there is an associated spontaneously broken global
(exact) symmetry~\cite{PhysRev.127.965}, and massive when an
approximate symmetry is broken~\cite{Weinberg:1972}. Both cases are
considered in this paper.

Two body kinematics dictate that the positrons in $\mu^+\to e^+X^0$ decay are
observable as a narrow peak at a momentum, $p_{X}$, determined by
the mass of the $X^0$ boson:
\begin{equation}\label{eq:px}
  p_e(m_{X}) = c\sqrt{\left(\frac{m_{\mu}^{2} - m_{X}^{2} +
        m_e^2}{2m_{\mu}}\right)^2 - m_e^2}
\end{equation} 
where $m_{\mu}$ is the mass of the muon, $m_e$ is the mass of the
positron, and $m_{X}$ is the mass of the boson generated by the LFV
process. 

This signal appears in addition to the three body positive muon
decay spectrum which, expressed in our measurement coordinates, is
\begin{equation}\label{eq:muenunu}
  \frac{d^{2}\Gamma}{dx d\cos\theta} =  
{\mathcal F}_{IS} (x; \rho, \eta) + P_{\mu}{\mathcal F}_{AS}(x;\xi,\delta)\cos\theta  
\end{equation}
where $\theta$\ is the angle of emission of the positron, $P_{\mu}$\
is the degree of polarization of the muon ensemble, and
$x = E_e/52.83 ~{\rm MeV}$ is the reduced positron
energy~\cite{TWIST:2011aa,Michel:1950,Fetscher:2010}. The measurement
z-axis points approximately opposite to the polarization direction, so
$P_{\mu} \sim -1$.  The muon decay parameters $\rho$, $\delta$, $\xi$,
and $\eta$ are bi-linear combinations of the weak coupling constants,
which assume values $\rho = \delta = 3/4, \xi=1, \rm{and }\ \eta=0$\
in the SM.

The  decay distribution of the positrons from the
$\mu^+\to e^+X^0$ process has an angular dependence
\begin{equation}\label{eq:mueJ}
  \frac{d\Gamma}{d\cos\theta} \propto 1 - A P_\mu \cos\theta
\end{equation}
We study the cases $A = 0$ (isotropic) and $A =
\pm1$ (maximally anisotropic). With this definition, $A=-1$\ 
corresponds to the asymmetry of the normal 3-body decay.  
Asymmetric two body muon decays are
predicted, for example, from Majoron production arising from a 
spontaneous violation of super-symmetric R-parity~\cite{Hirsch:2009ee}.

\section{The TWIST Experiment}

The TRIUMF Weak Interaction Symmetry Test (TWIST) has made an
order-of-magnitude improvement to the precision of the muon decay
parameters $\rho$, $\delta$, and
$P_{\mu}\xi$ \cite{Bayes:2011zza,TWIST:2011aa,Bueno:2011fq,Bueno:2012nr}. The
data, consisting of 1.1$\times 10^{10}$ stopped muon events, is appropriate for a
search for the inclusive two body decay. The experiment used highly
polarized muons delivered by the TRIUMF M13 beam line into a parallel
plane spectrometer immersed in a uniform 2 Tesla magnetic field. The
spectrometer consisted of 44 drift chambers (DCs) and 12 proportional
chambers (PCs) arranged symmetrically about a high purity metal stopping
foil. The stopping targets (75 $\mu$m Al or 30 $\mu$m Ag) also served
as the central PC cathode. The design and construction of this
detector has been described in detail
elsewhere~\cite{Henderson:2004zz}. The spectrometer was oriented so
that it had an approximate cylindrical symmetry centered on the muon
beam-line axis, which is then defined as the z-axis of the detector
coordinate system. It was constructed so that the position of the
detector elements, specifically the position of sense wires, is known
with a total precision of parts in $10^{5}$. The magnetic field was
mapped to a similar precision. These factors determine the absolute
momentum scale for particle trajectories measured in the detector.
 
\begin{figure}
  \centering
  \includegraphics[width=\columnwidth]{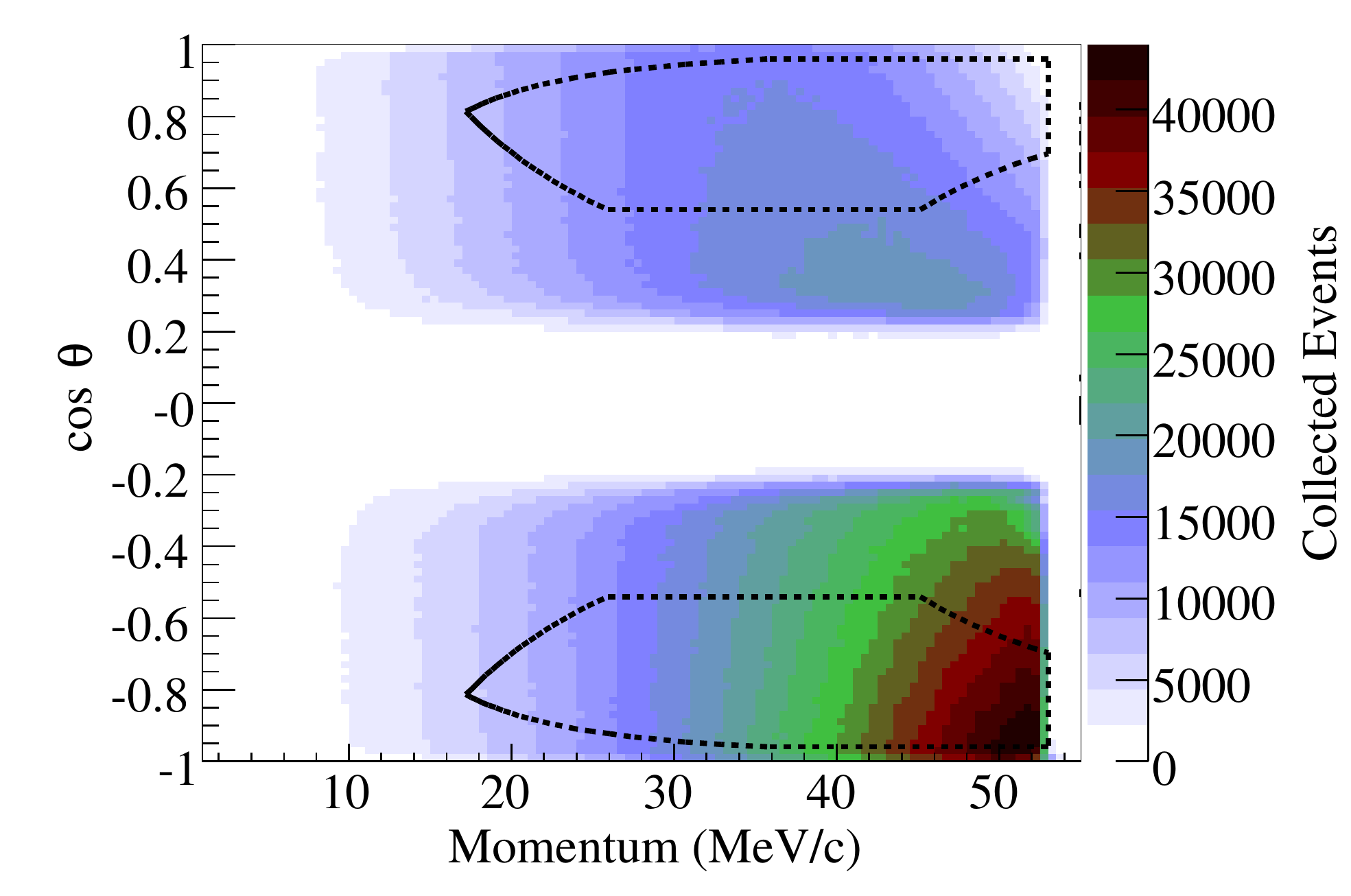}
  \caption{(color online) Spectrum of decay positron momenta and
    angles reconstructed from the TWIST spectrometer in bins of
    0.01$\times$0.5 MeV/c. Events used by this analysis are contained
    within the regions defined by the dashed line.}
  \label{fig:data_spectrum}
\end{figure}

Figure~\ref{fig:data_spectrum} shows our measured distribution of
positrons from muon decay binned by their total momentum $p_{tot} =
|\vec{p}|$ and $\cos\theta = p_{z}/p_{tot}$.  The planar geometry of the spectrometer allows for a
large angular acceptance of positrons resulting from decay in the
target foil, with a relatively simple momentum calibration. The
momentum resolution varies with $p_{tot}$ and $\theta$; at 52.8 MeV/c
the momentum resolution is (58
keV/c)/$|\sin\theta|$~\cite{TWIST:2011aa}.
 
Almost all of the physics data collected by the TWIST collaboration
during the 2006 and 2007 run periods were used for this two body decay
search. These data were subject to a sequence of event selection
criteria chosen to minimize the bias of comparisons between data and
simulation.  The event selection differs from the standard TWIST
analysis~\cite{Bayes:2011zza} only through the extension of the
momentum acceptance to include $p_{tot} < 53.0$ MeV/c. A total of
5.8$\times 10^{8}$ muon decay events were identified after the event
selection cuts were applied. The kinematic fiducial region has been
superimposed on the representative data spectrum shown in
Fig.~\ref{fig:data_spectrum}.

\section{Fitting Procedure}\label{s:Fitting}

The simulation of three body muon decay provides both the background
for the measurement and a model for the two body decay signal. The
response and acceptance of the detector are modelled using a detailed
GEANT 3.21 simulation.  A description of the simulation and its use in
the TWIST experiment is given in~\cite{rpm:2008prd,TWIST:2011aa}. We
use the simulation to generate the three body muon decay spectrum
$S_{M}$, which thus includes geometrical and physical effects, for use
in the fit to the data.  The periodicities in our plane and wire
spacing provide an example of such an effect because tracks at particular
momenta and angles may reconstruct with unusually large uncertainties.
Simulated muon decay events undergo the same reconstruction as the
standard data so these reconstruction inefficiencies are also included
in the simulated spectra.

The shape of the two body decay is presumed to be defined
by the momentum resolution of the reconstruction. The decay width may
only have a contribution if the lifetime of the $X^{0}$ boson produced in
the two body decay is less than $10^{-20}$~s. However, the analysis
will veto an event if a second charged particle appears in the
detector correlated to the decay at the stopping target. Based on the distance 
between the stopping target and the nearest DC, the lifetime of an $X^{0}$ boson allowed
by this analysis must be greater than 200 ps assuming that they decay
into  charged particles. Prompt $X^{0}$\ decays to $e^+e^-$\ are more 
strongly excluded from exclusive searches for such modes.

%The lifetime of the
%$X^{0}$ decay is similarly limited because the reconstruction and
%analysis requires that the decay positron produces a hit in the DC
%nearest to the stopping target or a minimum track length of 6~cm. Thus
%the lifetime of the $X^{0}$ must be greater than 200 ps.

The fit used to determine the branching ratio of the two body muon decay is
conducted through the use of a spectrum expansion originally developed
for the muon parameter fits in the standard TWIST
analysis~\cite{TWIST:2011aa}. An additional term is added to the fit
function based on Eq.~\ref{eq:muenunu}
\begin{eqnarray}
\nonumber S_{fit} = 
S_{M}(p_e, \cos\theta; \rho, \eta, \xi,\delta) \\
+  S_{A} ( m_{X} )B( m_{X}) \label{eq:mubr}
\end{eqnarray}
where $S_{A}(m_{X})$\ is the simulated two body decay positron
momentum distribution normalized to have the same integrated area as
the three body decay spectrum. Consequently the scale factor
${B}(m_{X}) = \Gamma(\mu^+\to e^+X^0)/\Gamma(\mu^+\to
e^+\nu_e\bar{\nu}_{\mu})$
is the branching ratio of a two body decay which produces a boson with
a mass $m_X$. Negative values of this scale factor are allowed by the
fit as deficits in the spectrum are statistically valid, but they
cannot correspond to physical particles.  The associated signal occurs
at a momentum, $p_{e}(m_X)$, defined in Eq.~\ref{eq:px}.

\begin{figure}[h]
  \begin{center}
    \includegraphics[width=1.05\columnwidth]{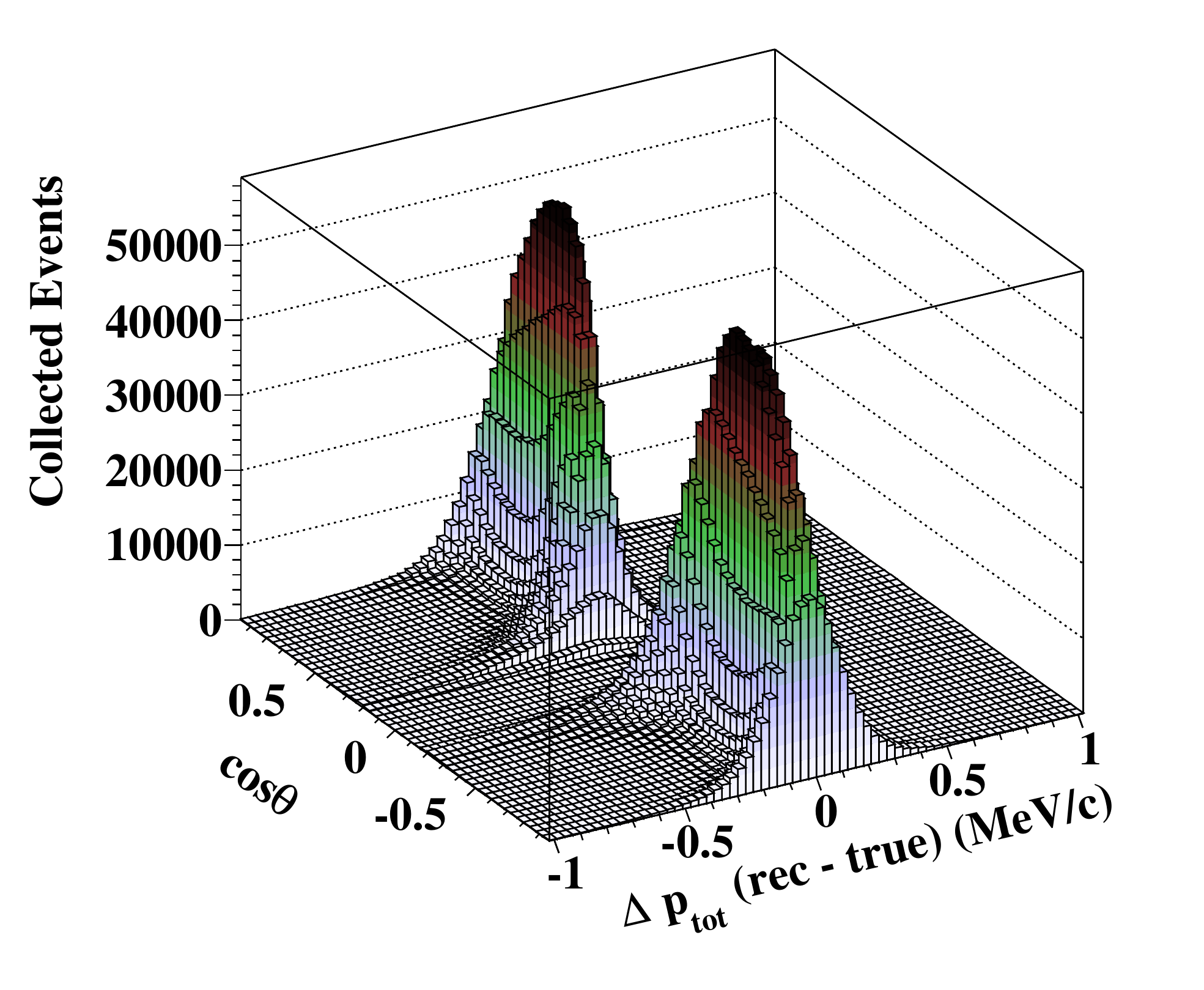}
  \end{center}
  \caption{(color online) The distribution of the momentum loss
    $\Delta p$ between the momentum of a simulated positron track at
    the time of decay and its momentum reconstructed from the positron
    track as a function of $\cos\theta$. This distribution, generated
    with 30~MeV/c$<p<$35~MeV/c is used to model a two body decay
    signal within that range after applying an offset in
    momentum. Similar distributions are derived for all other 5 MeV
    ranges. }
  \label{fig:pdist}
\end{figure}

The momentum distribution of positrons from two body decays was
derived from the difference of the reconstructed and true momentum,
$\Delta p = p_{\rm{rec}} - p_{\rm{true}}$, obtained as a function of
angle and momentum from the TWIST high statistics muon decay
simulations. Two-body decay distributions, $S_{A}(m_{X})$, were
generated for each of three cases tested: $A = -1$ with the same
anisotropy as the three body decay spectrum, $A = 0$ the isotropic
case, and $A = +1$\ where the anisotropy is opposite to that of the
three body decay spectrum.  The momentum response $S_{A}$\ of
positrons with momenta between 30 MeV/c and 35 MeV/c, which was used
to define the associated isotropic two body decay distribution for
this momentum range, is shown in Fig.~\ref{fig:pdist}. Two body decay
distributions  for other momentum ranges are similarly defined from
the momentum response defined from those ranges.

%The two body decay distribution has a width that varies with
%the positron momentum and angle in a way that is symmetric between the upstream
%and downstream spectra. The distribution of positrons from the two body
% decays was approximated from the difference of the reconstructed
%and true momentum, $\Delta p = p_{\rm{rec}} - p_{\rm{true}}$ of the
%TWIST high statistics muon decay simulations. 
%Two-body decay distributions, $\mathcal{F'}_{A}(m_{X})$, were generated
%for each of three cases tested:

To maximize the sensitivity to a narrow peak the data and simulation
are binned more finely than is optimum for the determination of the
muon decay parameters.  The branching ratio and decay parameters are
obtained from a $\chi^2$ fit of the data to Eq.~\ref{eq:mubr}. The fit
for the signal amplitude and the muon decay parameters $\rho$,
$P_{\mu}\xi\delta$, and $P_{\mu}\xi$ is performed for values of
$p_{e}(m_X)$ at 0.05 MeV/c intervals between 17.03 MeV/c and 52.83
MeV/c. This choice of interval size was made to limit running time of
the algorithm. The value of $\eta$ was fixed to
$-0.0036$~\cite{Danneberg:2005xv} in line with the TWIST muon decay
parameter analysis~\cite{Bayes:2011zza}. The decay parameters obtained
from these fits are consistent with those obtained when the two body
decay signal is omitted from the fit~\cite{TWIST:2011aa} at the level
of the measured statistical uncertainty, or a part in $10^{5}$, but
note that our results do not assume SM weak couplings.

The fitting procedure was assessed in two different ways. The first
applied the algorithm to a large number of statistically independent
simulations of three body muon decay spectra to study the statistical
distribution of peaks due to statistical fluctuations. The branching
ratios normalized by their uncertainties have a normal distribution
with a mean, $\mu = 0.01 \pm 0.03$ and a standard deviation,
$\sigma=0.98 \pm 0.03$, with a $\chi^{2}$ of 43 for 54 degrees of
freedom.

To assess the uncertainty introduced by the grid spacing used, two
body signals of a known amplitude were added to the three body decay
spectrum midway between the grid points.  A maximum deviation of 10\%
between the result of the fit and the signal amplitude was
found. Therefore we have increased by 10\% the upper limits of the
branching ratio obtained from our statistical analysis.

\section{Systematic Uncertainties for Endpoint Fits}

The momentum calibration used by the standard TWIST analysis distorts
the three body muon decay spectrum at the endpoint in a way that is
very similar to a two body decay signal. For this reason systematic
effects associated with the momentum calibration dominate the
uncertainty of a peak at the endpoint.  Two body decays with
$m_{X}<$~13~MeV/c$^2$, or less than 3 resolution widths from the edge
of the momentum spectrum at $\cos\theta = 0.8$, are not clearly
distinguishable from massless $X^0$ decays.

\begin{figure}[htbp]
\includegraphics[width=1.05\columnwidth]{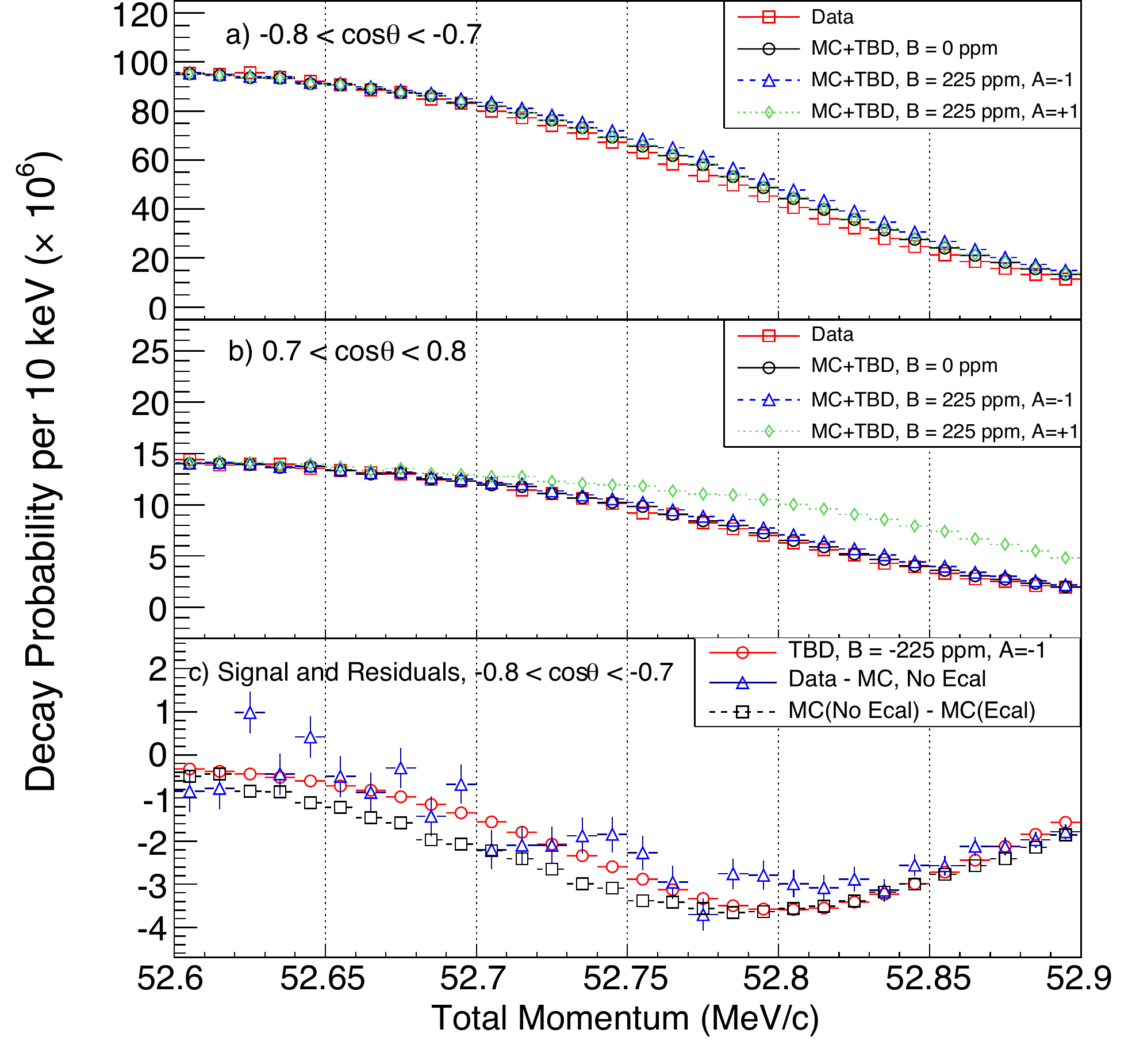}
\caption{(color online) Demonstration of the effect of two body decay
  (TBD) signals on the endpoint. Fig.~a(b) shows the decay probability
  for momentum near the endpoint of 
  upstream(downstream) spectra for data and Monte Carlo (MC)
  simulation with various two body decay signals added. The enhanced
  sensitivity to $A=1$ decays in the downstream spectrum is
  clear. Fig.~c shows a two body decay signal corresponding to a 
  branching ratio, $B$=-225~ppm, in comparison with the uncalibrated
  difference between data and MC (after the muon decay parameter fit,
  with no two body decay signal imposed) and the change in the shape
  of the spectrum produced by the momentum calibration (a 4.3 keV/c
  offset at $\cos\theta$ = -3/4), in the upstream spectrum as
  determined by the energy calibration fit.}
\label{fig:endpoint}
\end{figure}

In the standard TWIST analysis, a momentum calibration is performed by
matching the endpoint of the data spectrum to that of the simulated
spectrum assuming that any differences are linear with respect to
$\sec\theta$. The motivation for this procedure is to remove small
differences in the momentum of the reconstructed positron tracks in
data versus the simulation, consistent with differences of the energy
loss on the order of 10~keV/c. The differences between data and
simulation near the maximum possible momentum corresponding to
$E_{max}$ can be characterized by the momentum difference at
$|\cos\theta|= 3/4$, $\Delta p_{\pm3/4} = p_{\rm data} - p_{\rm sim}$
where $p_{\rm data}$, $p_{\rm sim}$ are the momenta at the spectrum
endpoint reconstructed from data and simulation. This sensitivity to
the momentum calibration is demonstrated in Fig.~\ref{fig:endpoint}
where the impact of two body decay signals of various anisotropies on
the endpoint muon decay spectrum is shown. Figure~\ref{fig:endpoint}(c)
explicitly compares a two body decay signal to the effect of a change
in the measured momentum calibration. In the analysis for light or
massless boson production the momentum calibration and its uncertainty
are obtained from known differences and uncertainties in the
simulation inputs, without using the end point calibration fits.

%These
%contributions and their impacts on the measurement of the $\mu^+\to
%e^+X^{O}$ branching ratio will be discussed in Section~\ref{s:sys}.

The corrections and uncertainties affecting the endpoint are
summarized in Table~\ref{tab:epschanges} for signals corresponding to
massless $X^0$ production. The offset and uncertainty in the spectrum
endpoint at $\cos\theta = -3/4$ indicates a magnitude for the
associated effect and is reported in the ``offset'' column of
Table~\ref{tab:epschanges}. The effects in the spectrum endpoint are
translated to uncertainties in the branching ratio using the
sensitivity of two body decay signals to variations in the momentum
calibration as shown in the right three columns. These sensitivities
are derived by altering the angle dependent and angle independent
components of the energy calibration, which are defined by a set of
four energy calibration parameters, and fitting for the
$\mu^{+}\to e^{+}X^{0}$ branching ratio for all accessible
$m_{X}$. Correlations between the endpoint calibration parameters are
included to reflect upstream/downstream and angle dependent
relationships for each contribution to the systematic
uncertainties. For example, a fit to an angle independent offset will
reveal that a 1 keV change will contribute 20 parts per million (ppm)
to the branching ratio at the endpoint and 0.2 ppm to branching ratios
for signals appearing at momenta less than 52 MeV/c assuming $A=1$.
The net effect of the uncertainties are generally much less than
this after correlations are included, as they are in
Table~\ref{tab:epschanges}.

The uncertainties in the stopping power of the detector materials and
the thickness of the muon stopping target produce a leading
contribution to biases and uncertainties in  $B(m_{X})$, as shown in
Table~\ref{tab:epschanges}. The momentum loss in the stopping target
alters the momentum offsets $\Delta p_{-3/4}$ and $\Delta p_{+3/4}$ by
the same amount, with a 100\% positive correlation between these
parameters. The measured difference in the muon stopping target
thickness from the value used in the simulation is 1.4$\pm$0.6 $\mu$m
for the silver target and 0.6$\pm$0.5 $\mu$m for the aluminum target.
The measurement was a destructive process conducted well after the
simulation was programmed and run. Averaging this effect over all data
sets yields a contribution of $-0.6\pm$0.4~keV/c to the momentum
offsets. Further uncertainties in the energy loss are associated with
the simulation of the target material which uses values taken from the
Berger-Seltzer report \cite{Berger:bk64}. In this case there is a 2\%
uncertainty in the calculated ionization energy loss and a 3\%
uncertainty in the radiative energy losses. In the detector stack,
events with large energy loss components are suppressed by the track
fitting procedure, which disassociates the trajectory into multiple
instances rather than changing the effective momentum of the fitted
helix. As a result, only the ionization energy loss uncertainties are
included for those materials.

The differences between the simulated and the true stopping position
of the muon introduces anti-correlated contributions to $\Delta
p_{-3/4}$ and $\Delta p_{+3/4}$. These were estimated to be 1.6 $\mu$m
in Ag and 3.8 $\mu$m in Al~\cite{TWIST:2011aa}. Averaging over all
data sets, this produces a change in the offsets of $\Delta
p_{-(+)3/4} = -(+)0.9\pm 1.0$ keV/c.

The space time relationship (STR) within the drift
cell~\cite{Grossheim:2010kk}, magnetic field, and detector dimension
uncertainties all affect the momentum offset and the angular
dependence of the endpoint. These systematic uncertainties are
independent upstream and downstream. A difference between data and
simulation of 1.4$\times10^{-4}$~T in the average magnetic field at
the position where it is monitored is predicted from a study of the
field mapping systematics\cite{TWIST:2011aa}. This alters the positron
momentum scale by
$-2.8\pm 1.5$~keV/c at $\cos\theta = -3/4$.  A fractional uncertainty
of 5$\times 10^{-5}$ in the detector length scale and thus the
position of the wire planes was calculated from the uncertainties of
the detector components. The uncertainty due to the STRs was estimated
from their difference when the STR for each wire plane is separately
determined from the data and when a plane-averaged STR is determined
from the simulation. There is negligible evidence of corrections due
to the STRs or a mis-calibration of the detector length scale.

\begin{table}
  \begin{tabular}{r|c|ccc}

    \hline\hline
    & Offset &  \multicolumn{3}{c}{Uncertainty in $B$ (in ppm)} \\
    Detector Property & (in keV/c) & $A=-1$&
    $A=0$&$A=+1$\\
    \hline
    Target Thickness & -0.6$\pm$0.4 & 7.6 & 2.5 & 0.8 \\
    Energy Loss in Target & 0.0$\pm$4.7 & 89.8 & 32.2 & 11.3 \\
    Stopping Distribution & -0.9$\pm$1.0 & 17.3 & 0.4 & 1.8 \\
    STRs & 0.0$\pm$3.1 & 49.1 & 10.8 & 4.3 \\
    Field Map Correction & -2.8$\pm$1.5 & 6.0 & 3.8 & 0.6 \\
    Detector Length & 0.0$\pm$4.3 & 12.9 & 9.3 & 0.9 \\
    Calibration Model & 0.0$\pm$1.6 & 21.8 & 8.1 & 2.3 \\
    Resolution &  0.0$\pm$3.0 & 21.4 & 7.6 & 3.1 \\
    \hline
    Total & -4.3$\pm$6.1 & 107.3 & 36.4 & 12.6 \\
    \hline\hline

  \end{tabular}
  \caption{Biases and uncertainties introduced to the momentum edge of
    the positron spectrum by various systematic effects. The endpoint
    offset is given as the change in the momentum edge at the center
    of the angular fiducial, $\cos\theta = -3/4$. The uncertainties
    in the offsets corresponding to each of these systematic effects
    produce the uncertainties in the two body decay branching ratios
    shown in the right three columns.}
  \label{tab:epschanges}
\end{table}

The above uncertainties assume a linear dependence of the momentum
calibration with respect to $\sec\theta$. However, the $\chi^{2}$
determined from the fits of the upstream momentum calibration exceeds
the number of degrees of freedom by a factor of 1.27, suggesting that
the model used to determine the energy calibration is not an ideal
model of the angular behavior at the endpoint. In absence of a
motivated correction to the model, an inflation of the statistical
uncertainty was introduced to account for this potential
uncertainty. The inflation of the uncertainty produces a systematic
bias in the endpoint momentum offset of 1.6 keV/c.

The contributions of each of these systematics to the value of the
endpoint offset is shown in Table~\ref{tab:epschanges}. The values of
$\Delta p_{+3/4} = -2.5 \pm 6.1$ keV/c and $\Delta p_{-3/4} = -4.3 \pm
6.1$ keV/c are consistent at the 1.5$\sigma$ level with the offset
obtained from fitting the endpoints of the data to the
simulation~\cite{TWIST:2011aa} that were used in the decay parameter
analysis.

The momentum resolution difference between data and simulation has an
upper limit of 3 keV/c based on the comparisons of fits to the
endpoint spectra using an error function convolved with a linear
approximation of the muon decay spectrum~\cite{TWIST:2011aa}. These
differences produce structure in the endpoint region that will alter
the two body decay fits. To evaluate the resolution sensitivity, the
simulation was smeared on an event-by-event basis by an additional 40
keV, which exaggerates the existing difference between data and
simulation by a factor of 3.55. A signal search was conducted on the
altered spectrum. The resulting uncertainties in $B$ were added to the
other uncertainties in quadrature to produce the total uncertainties
at each trial momentum. The resolution uncertainties obtained at
momenta less than 52 MeV/c are consistent with statistical noise as
expected. The contribution for massless decays is given in
Table~\ref{tab:epschanges}.

\section{Limits for Massive $X^O$\ Decays}

\begin{figure}[!t]
  \centering
    \includegraphics[width=\columnwidth]{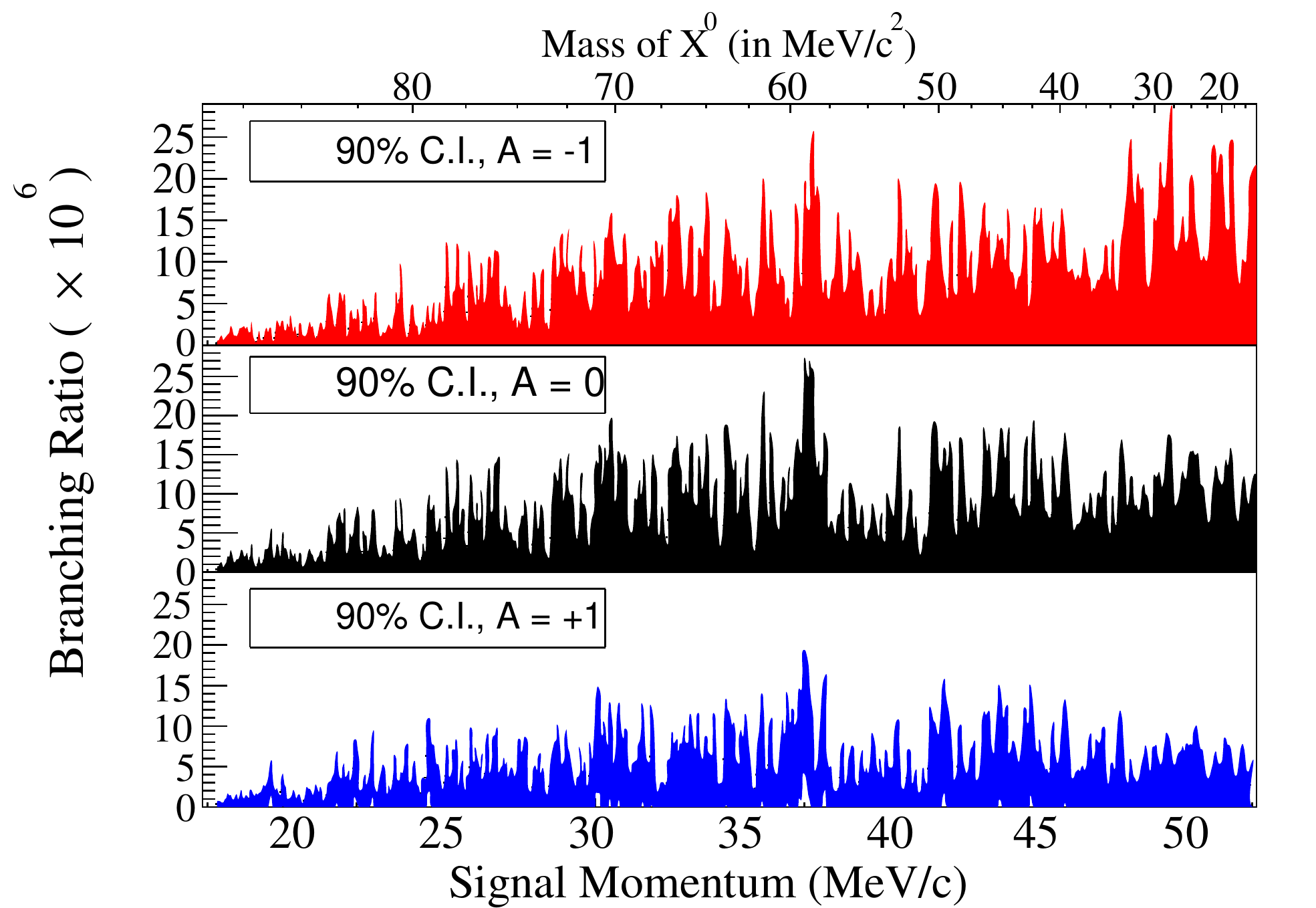}
  \caption{(color online) Confidence intervals set on branching ratios
    for $\mu^+ \to e^+ X^0$ decays determined from the muon decay
    spectrum for signals well separated from the
    endpoint. Statistical and energy calibration
    uncertainties are included.  }
\label{fig:brmass}
\end{figure}

The 90\% confidence intervals on $B(m_X)$ for $m_X>13$~MeV/c$^2$
are shown in Fig.~\ref{fig:brmass} for the three signal asymmetries.
These intervals were defined using the Feldman-Cousins (FC)
approach~\cite{Feldman:1998vn} and include both statistical and
systematic uncertainties. As expected from the number of $m_X$ grid
points on which the search is conducted, some of these lower limits
are non-zero. The significance (p-value) of these $B$ values is
assessed by calculating the probability that a peak with the same or
greater $B/\sigma$ will occur at any of the $m_X$ grid points due
to a random fluctuation.  This was obtained by running the two body
decay search on 1000 sets of randomized spectra and collecting the
most significant signal from each search. The randomized spectra were
generated by applying Poisson noise to the data and simulation. The
signal amplitudes measured from the randomized spectra less the
observed signal amplitude produces a probability distribution function
(PDF) consistent with the null hypothesis. The resulting PDF has an
appearance similar to a normal distribution and is used to define the 
p-value. Using the derived PDF is consistent with a
simpler approach to obtaining these p-values assuming normally
distributed uncertainties. These p-values, together with the average
limits obtained, are reported in Table~\ref{tab:signalresults}.  The
isotropic results can be compared directly to those of Balke \emph{et
  al.}\cite{Balke:1988by} and Bryman and
Clifford~\cite{Bryman:1986wn}.

\section{Limits for Massless $X^O$\ Decays}
  \begin{figure}[h]
    \includegraphics[width=\columnwidth]{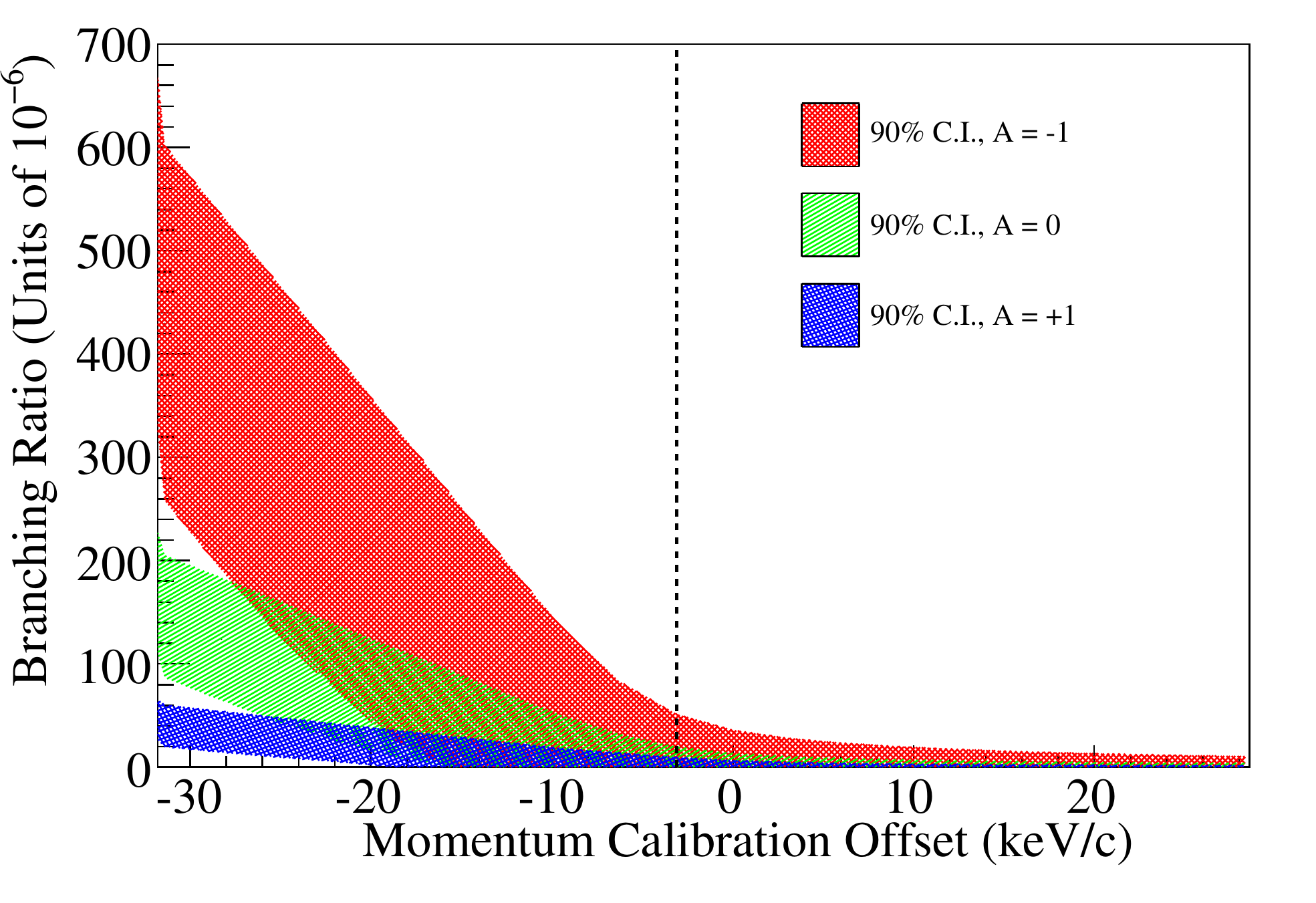}

  \caption{(color online) FC confidence intervals determined at the endpoint as a function of
    the momentum calibration offset. The black dotted line shows the best
    \emph{a priori} estimate of the momentum calibration as determined
    from Table~\ref{tab:epschanges}.}
    \label{fig:brmassless}
  \end{figure}
\begin{table}[t!]
  \centering
  \begin{tabular}{l|c|D{.}{.}{-1}|D{.}{.}{-1}}
    \hline\hline
    \multicolumn{2}{l|}{Decay Signal}&    \multicolumn{1}{c|}{90\%
    C.L.} & \multicolumn{1}{c}{p-value}\\
    \multicolumn{2}{l|}{} & \multicolumn{1}{c|}{(in ppm)} & \\
    \hline
    $A = 0$   &      Average	&       9   & \\
    & $p = 37.03$ MeV/c & 26 & 0.66 \\
    & Endpoint&     21 &     0.81 \\ \hline
    $A = -1$	&    Average	&   10  & \\
    & $p = 37.28$ MeV/c & 26 & 0.60 \\
    & Endpoint& 58 &	0.80  \\ \hline
    $A = +1$      &   Average	&    6  & \\
    & $p = 19.13$ MeV/c & 6 & 0.59 \\
    & Endpoint	&      10 &	0.90\\ \hline
    \hline
    \hline
    \multicolumn{4}{c}{Previous Results} \\
    \hline
    \multicolumn{2}{l|}{Balke \emph{et al.}~\cite{Balke:1988by}} &  100  &\\
    \multicolumn{2}{l|}{Bryman and Clifford~\cite{Bryman:1986wn}} &  300  &\\
    \multicolumn{2}{l|}{Jodidio \emph{et al.}~\cite{JodidioRHC}} &  2.6 & \\
    \hline\hline
  \end{tabular}
  \caption{The 90\% upper limits for the branching ratio of
    $\mu^+\to e^+X^0$ processes which
    produce positron signals with positive, negative, and no
    anisotropy. The average of the upper limits of $e^+$ signals
    produced in the presence of massive $X^0$ particles is shown for
    all three cases as well as similar limits associated with massless
    $X^0$ particles determined from the positron spectrum endpoint.
    The momentum, 90\% upper confidence limits, and p-value of
    the most significant massive signal is also given. The results 
    of Balke \emph{et al.} and Bryman and Clifford
    are directly comparable to the case of $\mu^+\to e^+X^0$ decays
    producing massive bosons with no anisotropy ($A = 0$), while the results
    of Jodidio are comparable to the production of massless $X^0$
    bosons, also assuming $A=0$. }
  \label{tab:signalresults}
\end{table}

The branching ratio limits quoted at the
endpoint and shown in Fig.~\ref{fig:brmassless} are based on the single fit of a $\mu^+\to e^+X^0$ signal at
$p_X = 52.83$ MeV/c. Values in  Table~\ref{tab:signalresults} use the momentum calibration calculated from
the systematic bias. All of the observed branching ratios are
consistent with statistical fluctuations.

The isotropic results can be compared directly to those of Jodidio
\emph{et al.}~\cite{JodidioRHC}. Those limits on a
$m_X=0$ signal are obtained from an accumulated 1.8 $\times 10^{7}$
muon triggers using a spectrometer with an angular acceptance such
that $\cos\theta > 0.975$. The three body muon decays are strongly
suppressed in this region. A consequence of the limited angular range
is a much larger muon sample density and effective sample size.  Since
the momentum resolution was also better than that of the TWIST
detector by a factor of 2 at similar angles, the upper limit on the
branching ratio is an order of magnitude smaller than the comparable
limits set by this work. However, the experiment was also insensitive
to signal anisotropies. Consequently, a signal with $A = -1$ would
have not been visible, while a signal with $A = +1$ would have been
excluded with a 1.3 ppm upper limit at 90\% confidence.

\section{ Conclusions}

No significant evidence for $\mu^+\to e^+X^0$\ decays has been found
in this search. The limits on these decays for 13 MeV$/c^2 < m_{X^0} <
80$ MeV$/c^2$, where the $X^0$ decay is not observed, have been
improved by a factor of 10 over previously published limits. The
dependence of these limits on the decay anisotropy has been studied
for the first time.

Due to the systematics associated with the detailed understanding of
the decay positron spectrum endpoint, our limits on $\mu^+\to e^+X^0$
processes with $m_X <13$\ MeV/$c^2$ are much less restrictive. For this
range we have reported the first inclusive limit on decays having the
same anisotropy as ordinary muon decay, while for other anisotropies
the Jodidio \emph{et al.} measurement is more sensitive.

\section{Acknowledgments}
All early TWIST collaborators and students deserve profound thanks for
their efforts in producing these results. Particular thanks go to
N.~Rodning, C.~Ballard, M.~Goyette, S.~Chan, A.~Rose, P.~Winslow, and
the TRIUMF cyclotron operations, beam lines, and support
personnel. This work was supported in part by the Natural Science and
Engineering Research Council and the National Research Council of
Canada, the Russian Ministry of Science, and the U.S. Department of
Energy. Computing resources were supplied by WestGrid and
Compute/Calcul Canada.

\bibliography{raredecays.bib}
\end{document}